\documentclass[conference, 10pt]{IEEEtran}
\IEEEoverridecommandlockouts
\usepackage[noadjust]{cite}
\usepackage{amsmath,amssymb,amsfonts}
\usepackage{algorithm}
\usepackage{algorithmic}
\usepackage{graphicx}
\usepackage{textcomp}
\usepackage{xcolor}
\usepackage{tikz}
\usepackage{subcaption}
\usepackage{mathtools}
\usepackage{lipsum}

\usetikzlibrary{shapes,arrows,shadows,positioning}
\usetikzlibrary{arrows, arrows.meta, shapes, shapes.multipart, trees, positioning,
	backgrounds, fit, matrix, petri, calc, automata, chains}
\usepackage{ellipsis}
\usetikzlibrary{calc}
\usetikzlibrary{decorations.pathreplacing,decorations.markings,shapes.geometric}
\usetikzlibrary{patterns}
\def\BibTeX{{\rm B\kern-.05em{\sc i\kern-.025em b}\kern-.08em
    T\kern-.1667em\lower.7ex\hbox{E}\kern-.125emX}}

\author{
	\IEEEauthorblockN{Onur Ayan\IEEEauthorrefmark{1}, H. Murat G\"ursu\IEEEauthorrefmark{1} Sandra Hirche\IEEEauthorrefmark{2}, Wolfgang Kellerer\IEEEauthorrefmark{1}} \\ \vspace*{-.4cm}
	\IEEEauthorblockA{\IEEEauthorrefmark{1}Chair of Communication Networks, Technical University of Munich}\\ \vspace*{-.4cm}
\IEEEauthorblockA{\IEEEauthorrefmark{2}Chair of Information-oriented Control, Technical University of Munich}\\ 
\vspace*{-.4cm}
\IEEEauthorblockA{\{onur.ayan, murat.guersu, hirche, wolfgang.kellerer\}@tum.de}
}
\title{AoI-based Finite Horizon Scheduling for Heterogeneous Networked Control Systems}

\definecolor{myred}{RGB}{220,43,25}
\definecolor{mygreen}{RGB}{0,146,64}
\definecolor{myblue}{RGB}{0,143,224}
\definecolor{mygray}{gray}{0.80}
\definecolor{mylightergray}{gray}{0.87}
\definecolor{mylightestgray}{gray}{0.95}

\usepackage{soul}

\newif\ifcomments

\commentstrue

\newcommand{\E}{\mathbb{E}}
\newcommand{\tr}{\mathsf{tr}}

\newcommand{\plant}{\mathcal{P}_i}
\newcommand{\sensor}{\mathcal{S}_i}
\newcommand{\controller}{\mathcal{C}_i}

\newcommand{\minus}{\scalebox{0.75}[1.0]{$\,-\,$}}
\newcommand{\floor}[1]{\left\lfloor #1 \right\rfloor}
\newcommand{\ceil}[1]{\left\lceil #1 \right\rceil}

\begin{document}

\maketitle

\begin{abstract}
Age of information (AoI) measures information freshness at the receiver. AoI may provide insights into quality of service in communication systems. For this reason, it has been used as a cross-layer metric for wireless communication protocols. In this work, we employ AoI to calculate penalty functions for a centralized resource scheduling problem. We consider a single wireless link shared by multiple, heterogeneous control systems where each sub-system has a time-varying packet loss probability. Sub-systems are competing for network resources to improve the accuracy of their remote estimation process. In order to cope with the dynamically changing conditions of the wireless link, we define a finite horizon age-penalty minimization problem and propose a scheduler that takes optimal decisions by looking $H$ slots into the future. The proposed algorithm has a worst-case complexity that grows exponentially with $H$. However, by narrowing down our search space within the constrained set of actions, we are able to decrease the complexity significantly without losing optimality. On the contrary, we show by simulations that the benefit of increasing $H$ w.r.t. remote state estimation performance diminishes after a certain $H$ value.
\end{abstract}


\section{Introduction}
\label{sec:intro}
The next generation communication networks are envisioned to support a wide range of machine-to-machine (M2M) and internet-of-things (IoT) applications, e.g., vehicular networks, industrial automation, smart grid. From system theoretical point of view, they can be modeled as \textit{networked control systems (NCS)} which are feedback control loops that are closed over a communication network. 

Studies show that the performance of NCS, i.e., quality of control (QoC), is tightly coupled with the service provided by the communication network. However, the metrics that have been extensively used to measure the performance of human-oriented networks, such as delay, packet loss and throughput, are not sufficient to capture the QoC requirements of heterogeneous NCS applications. Thus,
new cross-layer metrics have been adopted for communication
protocol design. 



Age of information (AoI) is such a relatively new metric that measures information freshness at the receiver monitoring a remote process \cite{kaul2012real} and used as a cross-layer metric in wireless medium access protocol design \cite{kadota2016minimizing, kadota2018scheduling, kosta2019age}. AoI is defined as the elapsed time since the generation of the freshest information available at the receiver. AoI increases linearly in time, and drops upon a reception of a new update. In such a setting, regardless of its cause, a low update rate at the receiver leads to staleness of the information. Therefore, AoI is a metric combining multiple metrics of communication networks such as delay and packet loss into a single one. This renders AoI easy to be used as an intermediate metric to calculate control system metrics.
In \cite{kadota2016minimizing}, authors consider a multi-user scenario in which each user is prone to a time-invariant packet loss. They formulate a discrete-time wireless scheduling problem and show analytically that the minimum AoI is achieved by updating the user with the highest AoI. They extend their findings in \cite{kadota2018scheduling} for the equivalent scenario and provide a complexity analysis of the proposed optimal scheduling algorithm together with other sub-optimal, low-complexity algorithms.


 AoI has been considered with linear and homogeneous dynamics in time and it cannot be mapped into closed loop control applications directly, where in general the state evolution is a nonlinear function of time and depends on the system parameters. Therefore, the introduction of non-linear aging concept in \cite{kosta2017age} facilitated the adoption of AoI in NCS research. In the context of NCS, AoI has first been used in \cite{ayan2019age} where authors employ AoI as an intermediate metric for a resource scheduling problem. They consider a lossless network with limited resources. They show that the proposed  heuristic scheduler achieves a higher QoC than the optimal AoI scheduler. The work in \cite{klugel2019aoi} follows the same AoI-based derivations from \cite{ayan2019age} and proposes a cost optimal, threshold-based transmission policy for a single-loop scenario.

The closest work to ours is \cite{ayan2019optimal}, which studies the problem of scheduling multiple, heterogeneous control systems sharing a common link with packet loss. Similar to the existing literature, they consider the packet loss to be constant over time. They employ the AoI-based metric, i.e., age-penalty, derived in \cite{ayan2019age} and propose an optimal scheduler that minimizes the discounted age-penalty over an infinite time horizon. However, this solution is not applicable to scenarios with changing channel conditions over time. 

To the best of our knowledge both in AoI and NCS research, there is no previous work that considers time-varying channel conditions and at the same time provides optimality in control-aware scheduling. It is a challenging task to guarantee optimality when the network conditions are changing over time, as the scheduling decisions remain optimal only for a certain amount of time. 

In this work, we aim to address the open research question of scheduling decisions to guarantee optimal NCS behavior for changing channel conditions. To this end, we study control dependent age-penalty minimization of feedback control loops that share a wireless communication link with time-varying packet loss probabilities. Time varying conditions in the network prevent the calculation of infinite horizon optimal scheduling policies. On contrary, {by applying stochastic optimization and dynamic programming,} we propose an online, centralized scheduling policy that is age-penalty optimal for a finite horizon $H$. The worst-case complexity of our approach grows exponentially but the performance gain expected in return diminishes after a certain $H$ value, hence a finite $H$ value can be selected without incurring too much complexity. To the best of our knowledge, this is the first work proposing an optimal scheduling algorithm for NCS with varying channel conditions. 

\section{System Model}
\label{sec:model}

We consider a communication network consisting of $N$ independent, linear time-invariant (LTI) control sub-systems. Each sub-system $i$ consists of a plant $\plant$, a sensor $\sensor$ and a controller $\controller$ that are connected through communication. The  state of $\plant$ is observed via an ideal link by $\sensor$ and transmitted in a packet over a wireless link to $\controller$. The wireless link is shared among all  $N$ sub-systems. We assume each controller-plant pair to be co-located and hence connected through an ideal link.  Moreover, each packet contains a single observation. 


Time is divided into slots which is also the smallest time unit in our model. We use $t \in \mathbb{N}$ to index time slots. Each packet transmission starting in slot $t$ ends within the same slot. Moreover, medium access is controlled by a centralized scheduler. That is, each sensor $\sensor$ transmits only when it has been allocated a slot by the scheduler. The remaining sensors that are granted access, do not transmit. Suppose an indicator variable $\delta_i(t) \in \{0, \, 1 \}$ that takes the value of 1 if the $i$-th sub-system is scheduled for transmission in slot $t$. Then, for any other sensor $j \neq i$, $\delta_{j}(t) = 0$ holds; or equivalently $\sum_{i=1}^{N} \delta_i(t) \leq 1$. Any successfully transmitted information in slot $t$ is available at $\controller$ at $t+1$.

We consider a time-varying link quality between each sensor-controller pair that affects the loss probability of transmissions. Each packet transmitted at $t$ is lost with probability $p_i(t)$ or received successfully with probability $1 \minus p_i(t)$. Given that $\sensor$ is scheduled, let $\gamma_i(t) = 1$ indicate a successful transmission with probability $\Pr\left[ \gamma_i(t) = 1 ~|~ \delta_i(t) = 1 \right] = 1 \minus p_i(t)$ and $\gamma_i(t) = 0$ a failed transmission with probability $\Pr\left[ \gamma_i(t) = 0 ~|~ \delta(t) = 1 \right] = p_i(t)$, respectively.

 We assume block fading where the link quality is approximately constant for a certain number of slots. This period, in which the link quality is assumed to be constant, is called \textit{coherence time}. The scheduler can measure the instantaneous link quality for each sub-system, hence is aware of the loss probabilities $p_i(t)$. However, the coherence time is unknown to the scheduler. 

The generation of packets at $\sensor$ is periodic with $D_{i} \in \mathbb{Z}^+$ slots between two consecutive observations. We call the generation of a packet a \textit{sampling event} and the time between two consecutive sampling events a \textit{sampling period}. Let $t_{i,o}$, selected uniformly in $[0, D_i)$, denote the time of the first sampling event of sub-system $i$\footnote{One can interpret $t_{i,o}$ as the time offset which initiates the operation of sub-systems. As we allow $t_{i,o}$ to be different for each sub-system, we do not enforce any synchronization of sampling among users. }. Thus, the set of slots in which a sampling event occurs is given as:
\begin{equation}
\mathcal{G}_i \triangleq \lbrace t_{i,o}, \; t_{i,o} + D_{i}, \; t_{i,o} + 2 D_{i}, \; \dots \rbrace.
\end{equation}

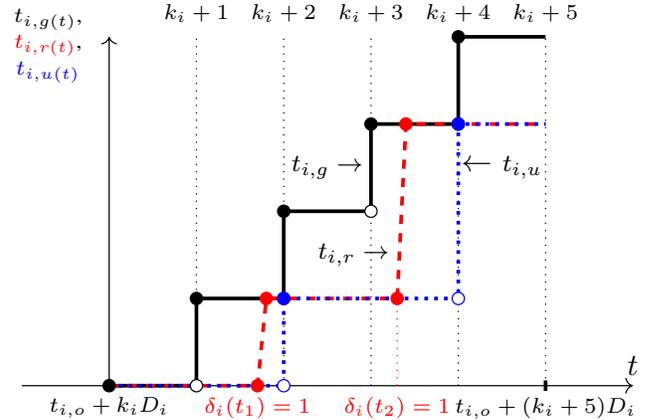
\begin{figure}[t]
	\centering
	\resizebox{\columnwidth}{!}{%
		\begin{tikzpicture}

\draw[->] (-1,0) -- (6,0) node[anchor=south] {$t$};
\draw	(0,4.5) node[anchor=east] {}
		(1,4.5) node[anchor=north] {\scriptsize$k_i+1$}
		(2,4.5) node[anchor=north] {\scriptsize$k_i+2$}
		(3,4.5) node[anchor=north] {\scriptsize$k_i+3$}
		(4,4.5) node[anchor=north] {\scriptsize$k_i+4$}
		(5,4.5) node[anchor=north] {\scriptsize$k_i+5$}
		(0,-0.0) node[anchor=north]
		{\scriptsize$t_{i,o} + k_i D_i$}
		(5,-0.0) node[anchor=north]
		{\scriptsize$t_{i,o}+(k_i+5)D_i$}
		(1.7,-0.0) node[anchor=north] {\color{red}\scriptsize$\delta_i(t_1)=1$}
		(3.3,-0.0) node[anchor=north] {\color{red}\scriptsize$\delta_i(t_2)=1$};

\node[] at (-0.7, 4.2) {\scriptsize$t_{i,g(t)}$,};
\node[] at (-0.7, 3.9) {\color{red}{\scriptsize$t_{i,r(t)}$}\color{black}{\scriptsize,}};
\node[] at (-0.7, 3.6) {\color{blue}{\scriptsize$t_{i,u(t)}$}};

\draw[very thick, black] (5,-0.06) --  (5,+0.06);
	
\draw[->] (0,0) -- (0,4) node[anchor=east] {};
\draw[dotted] (1,0) -- (1,4);
\draw[dotted] (2,0) -- (2,4);
\draw[dotted] (3,0) -- (3,4);
\draw[dotted] (4,0) -- (4,4);
\draw[dotted] (5,0) -- (5,4);
\draw[dotted,red] (3.3,0) -- (3.3,1);

\draw[very thick, black, solid] (0,0) --  (1,0) -- (1,1) -- (2,1) -- (2,2) -- (3,2) --  (3,3) -- (4,3) -- (4,4)-- (5,4);

\draw[very thick, red, dashed] (0,0) --  (1.7,0) -- (1.8,1) -- (3.3,1) -- (3.4,3) -- (5,3);

\draw[very thick, blue, dotted] (0,0) --  (2,0) -- (2,1) -- (4,1) -- (4,3) -- (5,3);

\draw (2.5, 2.5) node { \footnotesize$t_{i,g}\rightarrow$};

\draw (2.8, 1.5) node { \footnotesize$t_{i,r}\rightarrow$};

\draw (4.5, 2.5) node { $\leftarrow\,\,$\footnotesize$t_{i,u}$};

\node[circle,draw=black, fill=white, inner sep=0pt,minimum size=4pt] (b) at (1,0) {};
\node[circle,draw=black, fill=white, inner sep=0pt,minimum size=4pt] (b) at (2,1) {};
\node[circle,draw=black, fill=white, inner sep=0pt,minimum size=4pt] (b) at (3,2) {};
\node[circle,draw=black, fill=white, inner sep=0pt,minimum size=4pt] (b) at (4,3) {};

\node[circle,draw=black, fill=black, inner sep=0pt,minimum size=4pt] (b) at (0,0) {};
\node[circle,draw=black, fill=black, inner sep=0pt,minimum size=4pt] (b) at (1,1) {};
\node[circle,draw=black, fill=black, inner sep=0pt,minimum size=4pt] (b) at (2,2) {};
\node[circle,draw=black, fill=black, inner sep=0pt,minimum size=4pt] (b) at (3,3) {};
\node[circle,draw=black, fill=black, inner sep=0pt,minimum size=4pt] (b) at (4,4) {};

\node[circle,draw=blue, fill=white, inner sep=0pt,minimum size=4pt] (b) at (2,0) {};
\node[circle,draw=blue, fill=white, inner sep=0pt,minimum size=4pt] (b) at (4,1) {};

\node[circle,draw=blue, fill=blue, inner sep=0pt,minimum size=4pt] (b) at (2,1) {};
\node[circle,draw=blue, fill=blue, inner sep=0pt,minimum size=4pt] (b) at (4,3) {};

\node[circle,draw=red, fill=red, inner sep=0pt,minimum size=4pt] (b) at (1.7,0) {};
\node[circle,draw=red, fill=red, inner sep=0pt,minimum size=4pt] (b) at (3.3,1) {};

\node[circle,draw=red, fill=red, inner sep=0pt,minimum size=4pt] (b) at (1.8,1) {};
\node[circle,draw=red, fill=red, inner sep=0pt,minimum size=4pt] (b) at (3.4,3) {};

\end{tikzpicture}
	}
	\caption{Evolution of generation time $t_{i,g}(t)$, received time $t_{i,r}(t)$ and the update time $t_{i,u}(t)$ depicted in y-axis versus time in x-axis. $t_{i,g}(t)$ and  $t_{i,u}(t)$ are updated periodically with $D_i$ slots , while $t_{i,r}(t)$ can be updated asynchronously. On x-axis with $\delta_i(t_{1,2})=1$ two cases of successful packet transmission for sub-system $i$ are depicted.}
	\label{fig:timing_diagram}
		
\end{figure}

Under the assumption that status is Markovian, having received an update, the controller does not benefit from receiving an older observation. Thus, older packets are considered to be obsolete and ``non-informative''.  Therefore, $\sensor$ discards any older packet upon the arrival of a new update. 
We introduce a variable $t_{i,g}(t) \in  \mathcal{G}_i$ that denotes the generation time of the packet waiting for transmission. Note that, $t \in \mathcal{G}_i$ are the only slots, in which the information at $\sensor$ is updated. Hence, time dynamics of $t_{i,g}$ follows as:

\begin{equation}
	t_{i,g}(t+1) =
	\begin{cases}
	t+1 & \text{, if } t+1 \in \mathcal{G}_i \\
	t_{i,g}(t) & \text{, otherwise}	
	\end{cases}.
\end{equation}
similar to zero-order hold (ZOH) behavior to represent discrete time signals in continuous time.
An exemplary evolution of $t_{i,g}(t)$ is depicted in Fig.~\ref{fig:timing_diagram} where a discrete staircase behavior is observed. Additionally, the generation time of the freshest packet received by $\controller$ until time $t$ is denoted by $t_{i,r}(t)$ with: 
\begin{equation}
t_{i, r}(t+1) =
\begin{cases}
t_{i, g}(t) & \text{, if } \delta_i(t) \cdot \gamma_i(t) = 1 \\
t_{i, r}(t) & \text{, otherwise}	
\end{cases}.
\end{equation}
In Fig.~\ref{fig:timing_diagram} two successful updates demonstrate the relationship between $t_{i, r}(t+1)$ and $t_{i, g}(t)$. 

We consider the behavior of the $i$-th control sub-system is represented by the following LTI model in discrete time: 
\begin{equation}
\label{eq:discretemodel}
\boldsymbol{x}_i[k_i+1] = \boldsymbol{A}_i \boldsymbol{x}_i[k_i] + \boldsymbol{B}_i \boldsymbol{u}_i[k_i] + \boldsymbol{w}_i[k_i]
\end{equation}
with time-invariant system matrix $\boldsymbol{A}_i \in \mathbb{R}^{n_i \times n_i}$ and input matrix $\boldsymbol{B}_i \in \mathbb{R}^{n_i \times m_i}$. In addition, $\boldsymbol{w}_i \in \mathbb{R}^{n_i}$ represents the system noise characterized by a multi-variate Gaussian distribution with zero mean and covariance matrix $\boldsymbol{\Sigma}_i \in \mathbb{R}^{n_i \times n_i}$, i.e., $\boldsymbol{w}_i \sim \mathcal{N}(\boldsymbol{0}, \boldsymbol{\Sigma}_i)$. Time step, $k_i$, is a counter variable that is initialized with zero and increased by one after each sampling event. Therefore, it indicates in which sampling period a sub-system currently is. Since we assume the sub-systems to operate slower than the network, in slot granularity the evolution of $k_i$ is not strictly increasing. One can find the current $k_i$ of a sub-system at any slot $t$ simply by\footnote{For $t < t_{i,o}$, $k_i(t)$ takes a negative value. We allow this as $t_{i,o}$ defines the initialization of sub-systems and the system behavior before the operation starts is not taken into account for the remaining analysis.}: 
\begin{equation}
	\label{eq:k_map}
	k_i(t) = \floor{\frac{t - t_{i,o}}{D_i}}.
\end{equation}
Furthermore, $\boldsymbol{x}[k_i] \in \mathbb{R}^{n_i}$ and $\boldsymbol{u}[k_i] \in \mathbb{R}^{m_i}$ denote the system state and control input in sampling period $k_i$, respectively. At the beginning of each sampling period $k_i$, i.e., at $t = t_{i,o} + k_i D_i$, the controller $\controller$ calculates $\boldsymbol{u}_i[k_i] \in \mathbb{R}^{m_i}$ based on the available observation history. The obtained control input is applied to $\plant$ during that sampling period. Therefore, we assume that, any state update arriving after the calculation of $\boldsymbol{u}[k_i]$ is first utilized for the next input $\boldsymbol{u}_i[k_i+1]$.

The discrete-time model enables that control and communication progress with different time steps. Namely, in control, changes occur at sampling events
\footnote{We assume that the sampling period of each sub-system is selected small enough such that the changes between consecutive sampling events are negligible.}
. This is a well-established approach to model digital control systems in the literature \cite{astrom2008feedback}. Thereby, it can be argued that from the application's perspective the information of interest ages in discrete steps every $D_i$ slots. Hence, we define AoI as the number of sampling periods elapsed since the generation of the freshest information at $\controller$. As a result, the AoI in any slot $t$ is normalized by the sampling period and can be determined as:
\begin{equation}
\label{eq:aoi}
\Delta_i(t) = \ceil{\dfrac{t - t_{i,u}(t)}{D_{i}}}. 
\end{equation}

with:
\begin{equation}
\label{eq:t_u}
t_{i,u}(t+1) =
\begin{cases}
t_{i,r}(t+1) & \text{, if } t+1 \in \mathcal{G}_i \\
t_{i,u}(t) & \text{, otherwise}	
\end{cases}.
\end{equation}
where $t_{i,u}$ denotes the generation time of the latest utilized packet by $\controller$. The distinction between $t_{i,r}$ and $t_{i,u}$ is necessary since the controller receives packets during each sampling period that affect neither the control input nor the system state until the next sampling event occurs. In Fig.~\ref{fig:timing_diagram} the update delay of $t_{i, u}(t)$ following the $t_{i, r}(t)$ is demonstrated. $t_{i, u}(t)$ waits for the control system period to update its value. One can imagine the incoming information being queued until it is made use of at the end of the sampling period. 

It is important to point out that, as the packets are received at least one slot delayed, they can be utilized first to obtain the following control input. Thus, the minimum AoI in our system is one, i.e., $\Delta_i(t) \geq 1, \; \forall i, t$. 

%

\subsection{Remote Estimation and Control Law}
In order to overcome the shortcomings in the sensor-to-controller link, namely the packet loss and limited network resources, we consider an \textit{estimation-based controller} that estimates the plant state remotely. Suppose $\boldsymbol{x}_i[k_i - \Delta_i[k_i]]$ is the freshest information used by $\controller$ that is $\Delta_i[k_i]$ sampling periods old. As in \cite{ayan2019age}, it can easily be shown that the conditional expectation of the state which minimizes the mean squared estimation error, i.e.,  $\boldsymbol{\hat{x}}_i[k_i] = \E\left[\boldsymbol{x}_i[k_i] ~|~ \Delta_i[k_i], \, \boldsymbol{x}_i[k_i - \Delta_i[k_i] \right]$ can be obtained from:
\begin{equation}
\label{eq:estimatedstate}
	\boldsymbol{\hat{x}}_i[k_i] = \boldsymbol{A}_i^{\Delta_i[k_i]} \,  \boldsymbol{x}_i[k_i - \Delta_i[k_i]] + \sum_{q=1}^{\Delta_i[k_i]} \boldsymbol{A}_i^{q - 1} \, \boldsymbol{B}_i \, \boldsymbol{u}_i [k_i - q].
\end{equation}
Note that the \textit{estimation-based controller} has to keep track of the last $\Delta_i[k_i]$ control inputs. However, this does not imply any additional communication effort as this information is already present at $\controller$. We further assume that each $\controller$ is aware of the time-invariant system parameters $\boldsymbol{A}_i$, $\boldsymbol{B}_i$ and $\boldsymbol{\Sigma}_i$.

Moreover, from Eq.~\eqref{eq:discretemodel} and \eqref{eq:estimatedstate} we calculate the estimation error $\boldsymbol{e}_i(t)$ as:
\begin{equation}
\boldsymbol{e}_i[k_i] \triangleq \boldsymbol{x}_i[k_i] - \boldsymbol{\hat{x}}_i[k_i] = \sum_{q=1}^{\Delta_i[k_i]} \boldsymbol{A}_i^{q-1} \, \boldsymbol{w}_i[k_i - q].
\end{equation}
As a result, the expected mean squared error (MSE) at $\controller$ is expressed by:
\begin{align}
\label{eq:estimationerror}
\E \left[ \left(\boldsymbol{e}_i[k_i]\right)^T \, \boldsymbol{e}_i[k_i]\right] & = \sum_{r=1}^{\Delta_i[k_i] - 1} \tr \left( \left(\boldsymbol{A}_i^T\right)^r  \left(\boldsymbol{A}_i \right)^r \boldsymbol{\Sigma}_i \right) \\ \nonumber
& \triangleq g(\Delta_i[k_i]).
\end{align}
The derivations can be found in \cite{ayan2019age}. It is important to emphasize that the expected MSE is only a function of AoI and independent from the system state $\boldsymbol{x}_i[k_i]$. Here, we define the $g(\Delta[k_i])$ as a age-penalty function that maps AoI to MSE.



In addition to estimating the system state remotely, the controller determines the control input from control law as: 
\begin{equation}
\label{eq:controllaw}
\boldsymbol{u}_i[k_i] = - \boldsymbol{L}_i^* \,\boldsymbol{\hat{x}}_i[k_i].
\end{equation}
where $\boldsymbol{L}_i^* \in \mathbb{R}^{m_i \times n_i}$ denotes the optimal state feedback gain matrix. $\boldsymbol{L}^*_i$ is obtained from:
\begin{equation}
\label{eq:optimalgain}
\boldsymbol{L}_i^* = \left(\boldsymbol{R}_i + \boldsymbol{B}_i^T \boldsymbol{P}_i \boldsymbol{B}_i \right)^{-1} \boldsymbol{B}_i^T \boldsymbol{P}_i \boldsymbol{A}_i,
\end{equation}
which solves the discrete time algebraic Riccati equation:
\begin{equation}
\label{eq:riccati}
\boldsymbol{P}_i = \boldsymbol{Q}_{i} + \boldsymbol{A}_i^T \left(\boldsymbol{P}_i - \boldsymbol{P}_i \boldsymbol{B}_i ( \boldsymbol{R}_{i} + \boldsymbol{B}_i^T \boldsymbol{P}_i \boldsymbol{B}_i)^{- 1} \boldsymbol{B}_i^T \boldsymbol{P}_i \right) \boldsymbol{A}_i.
\end{equation}
$\boldsymbol{Q}_{i}$ and $\boldsymbol{R}_{i}$ are weighting matrices of appropriate size that penalize the state and control inputs in the infinite horizon, linear-quadratic-Gaussian (LQG) cost function $F_i$:
\begin{equation}
	F_i = \dfrac{1}{K} \limsup_{K \rightarrow \infty} \sum_{k_i=0}^{K-1} (\boldsymbol{x}_i[k_i])^T \boldsymbol{Q}_i \boldsymbol{x}_i[k_i] +  (\boldsymbol{u}_i[k_i])^T \boldsymbol{R}_i \boldsymbol{u}_i[k_i]. 
\end{equation}
$F_i$ is an indicator of control performance. The lower $F_i$ is, the higher is the \textit{quality of control (QoC)}. From Eq.~\eqref{eq:optimalgain} it is evident that the controller gain $\boldsymbol{L}_i^*$ depends only on constant system parameters and thus not optimized to take communication parameters into account. Communication-aware controllers exist in the literature, however they are beyond the scope of this paper.

\subsection{Problem Statement}
In  Eq.~\eqref{eq:estimationerror}, AoI appears as the only time-varying parameter. Thus, by controlling the AoI of sub-systems through scheduling, one can control the expected estimation performance. Our goal is to propose a scheduler that maximizes the overall estimation accuracy  in the network by minimizing the expected MSE. We are interested in class of scheduling policies $\pi$ that consist of a sequence of scheduling decisions $\mu(t')$ for the next $H \in \mathbb{Z}^+$ slots, i.e., $\pi= \{ \mu(t), \, \dots, \, \mu(t + H-1) \}$. $H$ is the \textit{finite horizon} parameter that defines how many future slots is being considered, while taking the current decision. Therefore, the selection of parameter $H$ controls the ``farsightedness'' of the adopted scheduler.

Each scheduling decision $\mu(t)$ defines the sub-system that is granted medium access where $\mu(t) = i$ and $\mu(t) = \varnothing$ implies $\delta_i(t) = 1$. First, let us define the network state $\boldsymbol{s} \in \mathbb{Z}^{3\cdot N}$ as:
\begin{equation}
\boldsymbol{s}(t) \triangleq \left[\boldsymbol{t}_g(t) \quad \boldsymbol{t}_r(t) \quad \boldsymbol{t}_u(t) \right]^T,
\end{equation} 
with:
\begin{align}
\boldsymbol{t}_g(t) &\triangleq \left[ t_{1,g}(t) ~ \dots ~ t_{N,g}(t) \right]^T ,\\
\boldsymbol{t}_r(t) &\triangleq \left[ t_{1,r}(t) ~ \dots ~ t_{N,r}(t) \right]^T ,\\
\boldsymbol{t}_u(t) &\triangleq \left[ t_{1,u}(t) ~ \dots ~ t_{N,u}(t) \right]^T.
\end{align}
Next, the set of possible actions given the network state $\boldsymbol{s}(t)$ is defined as:
\begin{equation}
\label{eq:admissibleactions}
\mathcal{M}(\boldsymbol{s}(t)) = \{\varnothing\} \cup \{i : t_{i,g}(t) > t_{i,r}(t) \}.
\end{equation}

Eq.~\eqref{eq:admissibleactions} implies that we constrain the action set and allow sub-systems only with a new information to transmit. Given a state $\boldsymbol{s}(t)$ and a scheduling decision $\mu(t) = i$, the transition probability to a possible next state $\boldsymbol{s}'$ strictly depends on the current condition of the corresponding link and is a function of $p_i(t)$. In particular, if $\boldsymbol{s}'$ is a possible next state after $\boldsymbol{s}(t)$, then $\Pr[\boldsymbol{s}(t+1) = \boldsymbol{s}' ~|~ \mu(t) = i, \boldsymbol{s}(t)  ] \in \{p_i(t), \, 1 - p_i(t)\}$.

We call the class of scheduling policies mapping states $\boldsymbol{s}(t)$ into possible scheduling decisions $\mu(t) \in \mathcal{M}(\boldsymbol{s}(t))$ \textit{admissible} and denote it by $\Pi$. Having the class of admissible policies, we can then confine our search within this class of narrowed down set. This enables us to reduce the search space and thereby complexity without losing optimality. 
Suppose $C(\boldsymbol{s}(t))$ denotes the total expected MSE over all sub-systems given the network state at time $t$ as:
\begin{equation}
\label{eq:gfunction}
C(\boldsymbol{s}(t)) =  \sum_{i=1}^{N}  g(\Delta_i[k_i(t)]).
\end{equation} 
Next, we employ $C(\boldsymbol{s}(t))$ as the state cost and define the expected finite horizon cost $J_\pi(\boldsymbol{s}(t))$ for any initial state $\boldsymbol{s}(t)$ and horizon $H$ as:
\begin{equation}
\label{eq:horizoncost}	
J_\pi(\boldsymbol{s}(t)) \triangleq \E_\pi \left[ \sum_{t' = t }^{t + H} C(\boldsymbol{s}(t'))     \right].
\end{equation}
Note the subscript $\pi$ in $J_\pi$ and $\E_\pi$ which indicates that Eq.~\eqref{eq:horizoncost} gives the expected cost when the scheduling policy $\pi = \{ \mu(t), \, \dots, \, \mu(t + H-1) \}$ is employed over the horizon $H$. 
The optimal policy $\pi^*$, in which we are interested, is the one that minimizes $J_\pi(\boldsymbol{s}(t))$; i.e.: 
\begin{equation}
\label{eq:minimizationproblem}
	J_{\pi^*}(\boldsymbol{s}(t)) = \min_{\pi \in \Pi} J_\pi (\boldsymbol{s}(t)) = J^*(\boldsymbol{s}(t)). 
\end{equation}
Throughout the remaining sections, we refer to the minimization problem in Eq.~\eqref{eq:minimizationproblem} as the $H$-stage problem and drop the subscript $\pi$ for brevity.
\section{Finite Horizon Scheduler}

We propose a \textit{finite horizon (FH) scheduler} that takes optimal scheduling decisions for the next $H$ transmission slots. It is shown in  \cite[p.~25]{bertsekas1995dynamic} that for the $H$-stage problem starting at state $\boldsymbol{s}(t)$ and time $t$, the optimal cost $J^*(\boldsymbol{s}(t))$ can be obtained by minimizing the right side of the Eq.~\eqref{eq:minimizationproblem} at each stage as: 

\begin{equation} 
\label{eq:dynamicprogramming}
	J_{t'}(\boldsymbol{s}(t')) = \min_{\mu(t') \in \mathcal{M}(t')} \E \left[ C(\boldsymbol{s}(t')) + J_{t' + 1}(\boldsymbol{s}(t' + 1))\right],
\end{equation} 
with the terminal cost:
\begin{equation}
\label{eq:terminalcost}
	J_{t + H}(\boldsymbol{s}(t + H)) = C(\boldsymbol{s}(t + H)). 
\end{equation}
The expectation is taken with respect to the total MSE in the network as defined in Eq.~\eqref{eq:estimationerror} and Eq.~\eqref{eq:gfunction}. Otherwise speaking, one can find the optimal cost $J^*(\boldsymbol{s}(t))$ by iterating backwards in time from stage $H \minus 1$ to stage $0$. Consequently, if at each iteration stage, $t'$, the optimal $\mu^*(t')$ action is taken, which minimizes Eq.~$\eqref{eq:dynamicprogramming}$, the policy $\pi = \{ \mu^*(t), \, \dots, \, \mu^*(t + H-1) \}$ is optimal achieving the minimum cost $J_{\pi ^*}(\boldsymbol{s}(t)) =  J_t(\boldsymbol{s}(t))$. The proof can be found in \cite{bertsekas1995dynamic}.

The $H$-stage problem, as described above, can be modeled as a tree structure, where each node inside the tree represents a network state $\boldsymbol{s}(t')$ occurring at time $t'$ with $t \leq t' \leq t + H$. The root of the tree corresponds to the current network state $\boldsymbol{s}(t)$. Moreover, nodes occurring in the same slot form a level together, where the root node constitutes the $0$-th level of the tree. Nodes in the last, i.e., $H$-th, are called \textit{leaf nodes}.

The backwards iteration to solve the $H$-stage problem can be viewed as visiting all levels of the tree starting from the leaf nodes and taking the optimal action at each level that minimizes the expected cost in the next level. Each node is assigned with a cost obtained from Eq.~\eqref{eq:dynamicprogramming}. As a result, the operation of the FH scheduler consists of the following steps:
\begin{enumerate}
	\item Initialize the current state $\boldsymbol{s}(t)$ as the root of a tree structure.
	
	\item Starting from the root, determine the possible actions at each node, i.e., $\mathcal{M}(t')$ for $\{t': t \leq t' < t + H \}$ and subsequently all possible next states $\boldsymbol{s}(t' + 1)$ when action $\mu(t')$ is taken.
	
	\item Add all possible next states as child nodes to the next level of the tree with the corresponding transition probabilities from the parent node.
	
	\item Repeat steps (2)-(4) until the $H$-th level of the tree is constructed.
	
	\item Assign costs to all states starting from the leaf nodes as in Eq.~\eqref{eq:dynamicprogramming} and \eqref{eq:terminalcost}.
\end{enumerate}

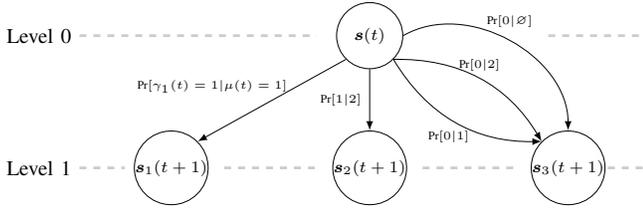
\begin{figure}[t]
	\centering
	\resizebox{\columnwidth}{!}{%
		\begin{tikzpicture}[>=latex]
\def\nodesize{1cm}

\node (s0) at (0,0) [draw, circle, align=center, inner sep=0pt, minimum size=\nodesize] {\scriptsize$\boldsymbol{s}(t)$};

\node (s1) at (-3,-2) [draw, circle, align=center, inner sep=0pt, minimum size=\nodesize] {\scriptsize$\boldsymbol{s}_1(t+1)$};

\node (s2) at (0,-2) [draw, circle, align=center, inner sep=0pt, minimum size=1cm] {\scriptsize$\boldsymbol{s}_2(t+1)$};
\node (s3) at (3,-2) [draw, circle, align=center, inner sep=0pt, minimum size=1cm] {\scriptsize$\boldsymbol{s}_3(t+1)$};

\draw[->] (s0.south west) [] to node [above, pos=0.5, xshift=-.9cm] (aa) {\tiny$\text{Pr}{[\gamma_{1}(t)=1|\mu(t)=1]}$} (s1.north east);
\draw[->] (s0.south) [] to node [pos=0.5, left] (bb) {\tiny$\text{Pr}{[1|2]}$} (s2.north);
\draw[->, bend right] (s0.south east) [] to node [pos=0.5, below, xshift=-.1cm] (bb) {\tiny$\text{Pr}{[0|1]}$} (s3.north west);
\draw[->, bend left] (s0.south east) [] to node [pos=0.5, above, yshift=0cm] (cc) {\tiny$\text{Pr}{[0|2]}$} (s3.north west);
\draw[->, bend left=60] (s0.east) [] to node [pos=0.5, above, yshift=0.1cm] (cc) {\tiny$\text{Pr}{[0|\varnothing] }$} (s3.north);

\node (lev0) [] at (-5,0) {\small Level 0};
\draw [dashed,mygray, very thick] (lev0.east) to node [] (xyz) {} (-.8,0);
\draw [dashed,mygray, very thick] (2.7,0) to node [] (xyz) {} (4.2,0);

\node (lev1) [] at (-5,-2) {\small Level 1};

\draw [dashed, mygray,very thick] (lev1.east) to node [] (xyz) {} (-3.7,-2);
\draw [dashed,mygray,very thick] (-2.2,-2) to node [] (xyz) {} (-.8,-2);
\draw [dashed,mygray,very thick] (.8,-2) to node [] (xyz) {} (2.2,-2);
\draw [dashed,mygray,very thick] (3.6,-2) to node [] (xyz) {} (4.2,-2);
%
%
\end{tikzpicture}
	}
	\caption{Example $1$ level tree structure with $2$ sub-systems, i.e., $H=1$ and $N=2$. Each edge is labeled with the corresponding transition probability, i.e., $\text{Pr}{[\gamma_{\mu(t)}(t)~|~\mu(t)]}$ as the conditional probability the outcome of the scheduled sub-system given the scheduling decision. The state $\textbf{s}_1$ and $\textbf{s}_2$ represent the cases where scheduled sub-system successfully updated the information in the controller while the $\textbf{s}_3$ represent the failure case where none of the sub-system can update the information.}
	\label{fig:tree}
\end{figure}

For better understanding, consider a toy example with $N = 2$. Suppose sub-systems are sampled every slot, i.e., $D_1 = D_2 = 1$, which is a special case and implies $t_{i,g}(t) = t, \, \forall \, i,t $. 
Fig. \ref{fig:tree} illustrates the first level of the tree structure constructed by the FH scheduler. Note that the tree has the root $\boldsymbol{s}(t)$. Outgoing edges are labeled with corresponding transition probabilities. They are written as conditional probabilities, that is, probability of the transmission outcome given the scheduling decision, i.e., $\text{Pr}{[\gamma_{\mu(t)}(t)~|~\mu(t)]}$.  Thus, the transition probabilities for sub-system $i$ can be written as: $\text{Pr}{[\gamma_{i}(t)=1~|~\mu(t)=i]} = 1 \minus p_i(t)$ and $\text{Pr}{[\gamma_{i}(t)=0~|~\mu(t)=i]} = p_1(t)$ where the case of no scheduled subsystem is a deterministic state of  $\text{Pr}{[\gamma_{1,2}(t)=0~|~\mu(t)=\varnothing]} = 1$. Consequently, given the initial state, $\textbf{s}(t) = [t \quad t \quad a \quad b \quad c \quad d]^T$, one can construct the next states at time $t+1$ as follows:
\begin{align*}
	s_{1}(t+1) &= [t+1 \quad t+1 \quad t \quad b \quad t \quad b]^T ,\\
	s_2(t+1) &= [t+1 \quad t+1 \quad a \quad t \quad a \quad t]^T ,\\
	s_3(t+1) &= [t+1 \quad t+1 \quad a \quad b \quad a \quad b]^T .
\end{align*}  
For the sub-system with a successful transmission, the $t_{i,r}(t+1)$ and $t_{i,u}(t+1)$ is set to $t_{i,g}(t)=t$ as given in the initial state $\boldsymbol{s}(t)$. For those sub-systems with a failed transmission or that are not scheduled, the $t_{i,r}(t+1)$ is kept same as $t_{i,r}(t)$  and $t_{i,u}(t+1)$ is updated to $t_{i,r}(t + 2)$.

Once the complete tree is constructed in a similar fashion, the FH scheduler assigns the cost and best action to each node as defined in Eq.~\eqref{eq:dynamicprogramming}. As a result, the optimal decision, $\mu^*(t)$, at current state $\boldsymbol{s}(t)$ is obtained and executed node and executes the optimal decision $\mu^*(t)$ at level $0$. It is important to emphasize that, even though the admissible and optimal policy $\pi^*$ consists of $H$ optimal actions, the steps (1)-(6) have to be repeated every time slot, as the root of the tree changes and only the action at level $0$ is extracted from the policy.

The complexity of the FH scheduling algorithm is dominated by the construction of the tree structure. If all sub-systems in the network have a sampling period of $1$ slot, which is the worst-case scenario in terms of complexity, the number of states in the tree equals to $(N+1)^{H+1} - 1$. Depending on the implementation, namely how often each node is visited to obtain the optimal action, the complexity becomes $c\cdot ((N+1)^{H+1} - 1)$, where c is a positive constant. Nevertheless, in the simplified Big-O notation, $c$ can be omitted and the complexity can be expressed as $\mathcal{O}(N^H)$. Increasing the sampling period further reduces the number of nodes. In particular, not having any update packet to transmit after a successful transmission, as packets are generated only at sampling events, reduces the number of possible actions at some states. In other words, if $t_{i,g}(t) = t_{i,r}(t)$, then $i \not \in \mathcal{M}(t)$. Thereby, the amount of nodes in the next level is reduced.
\section{Evaluation}
\label{sec:evaluation}
In order to evaluate the performance of our proposed FH scheduler, we set up a simulation environment with $N=3$ sub-systems as a minimal scenario to demonstrate the benefit of finite horizon minimization. We consider each sub-system to have a different system matrices, i.e., ${A}_{1,2,3} = \{1.0, \, 1.25, \, 1.5\}$, representing different task criticalities. The input matrix is given by $B_i = 1.0, \; \forall i$. Note that, $A_3 = 1.5$ represents the most challenging application. The control sub-systems start with $x_i[0] = w_i[0]$ for all $i$ and system noise is given by $w_i[k] \sim \mathcal{N}(0, 1)$, i.e., $\Sigma_i = 1$. Sub-systems are sampled periodically every 3 slots, i.e., $D_i =3$. The arrival time of the first sampling event is selected randomly from a discrete uniform distribution as $t_{i,o} = \mathcal{U}\{0, D_i - 1\}$ such that the sub-systems operate in a non-synchronized fashion.

We assume $\boldsymbol{Q}_i = 1$ and $\boldsymbol{R}_i = 0,$ $\forall i$ . That is, we take only the state cost into account but neglect the penalty for control effort. Therefore, from Eq.\eqref{eq:optimalgain} the optimal feedback gain matrix is determined as $\boldsymbol{L}_i^* = A_i$, which corresponds to deadbeat control strategy.

In order to model the time varying nature of the channel, we assume the loss probabilities, i.e., $p_i(t)$, to be normally distributed with mean $0.3$ and standard deviation $0.2$\footnote{The distribution is actually a rectified Gaussian distribution to reflect the possible loss values, i.e., all values below zero and above one are set to $0$ and $1$, respectively. This is aimed to mimic the packet loss performance in reference 3GPP channel model, where below and above a certain BER just a loss or success is observed, respectively.}. The coherence time of the channel, i.e., period of a constant $p_i(t)$, is $30$ slots. In other words, the loss probabilities do not vary for $30$ slots. The finite horizon scheduler is agnostic to the coherence time and uses $p_i(t)$ available at that time instance. 

We vary the finite horizon $H= \{1, \, \dots , \, 10\}$ to investigate the impact of making the FH scheduler more far- or shortsighted on the estimation performance. Each configuration is simulated for a duration of $D= 20 \, 000$ time slots and repeated $R = 200$ times. We measure the average mean squared error of a sub-system $i$ at each slot, i.e., $MSE_i = \frac{1}{D} \sum_{t = 1}^{D}  (\boldsymbol{e}_i(t))^T \boldsymbol{e}_i(t)$.
Additionally, the average MSE in the network, i.e., $\overline{MSE}$, is determined as $\overline{MSE} = \frac{1}{N} \sum_{i=1}^{N} MSE_i$.

\begin{figure}[t]
	\centering
	\includegraphics[width=\columnwidth]{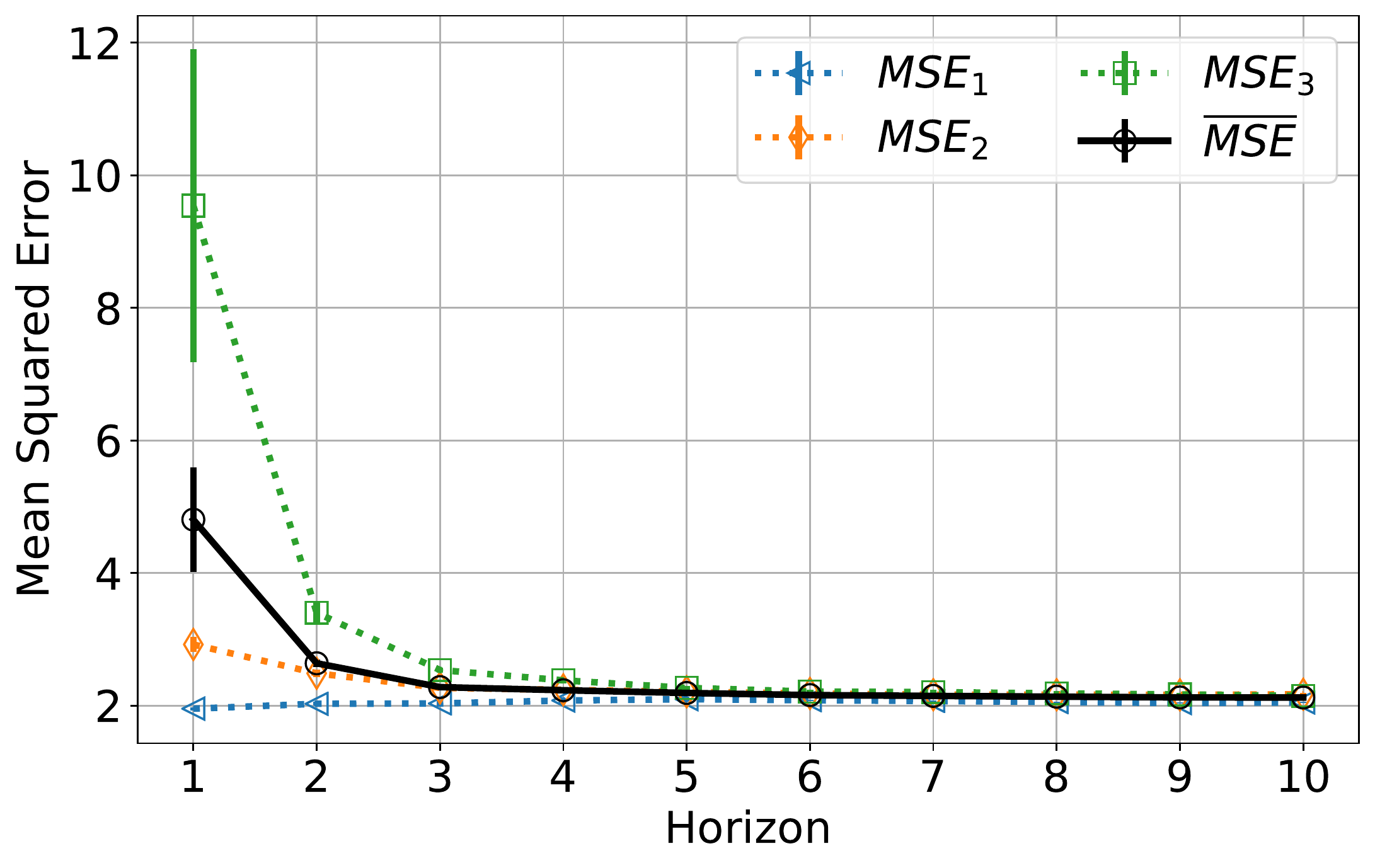}
	\caption{Solid line $\overline{MSE}$ illustrates the average MSE in the network per time slot. Dashed lines $MSE_1$, $MSE_2$ and $MSE_3$ show the the average MSE per time slot of each class with $A_1=1.0$, $A_2=1.25$ and $A_3=1.5$, respectively. Vertical bars represent $95\%$ confidence interval.}
	\label{fig:MSEavg}
\end{figure}

Fig.~\ref{fig:MSEavg} shows the estimation performance achieved by FH scheduler for varying $H$. We observe a drastic reduction of $\overline{MSE}$ as we increase the horizon from $H=1$ to $H=2$. However, beyond $H=2$, as the scheduler gets more farsighted, the performance gain diminishes. It is important to mention that for higher $H$ values, the averages of individual sub-systems, in other words the $MSE_i$ lines, tend to meet. This effect follows from equal weighting of MSE among sub-systems while obtaining the cost in Eq.~\eqref{eq:gfunction} which in turn leads to equal long-term averages as the scheduling actions become more ``long-term optimal''. In addition to MSE, we measure the AoI performance in the network similarly, i.e., $\Delta_i = \frac{1}{D} \sum_{t = 1}^{D}  \Delta_i(t)$ and $\overline{\Delta} = \frac{1}{N} \sum_{i=1}^{N} \Delta_i$.

\begin{figure}[t]
	\centering
	\includegraphics[width=\columnwidth]{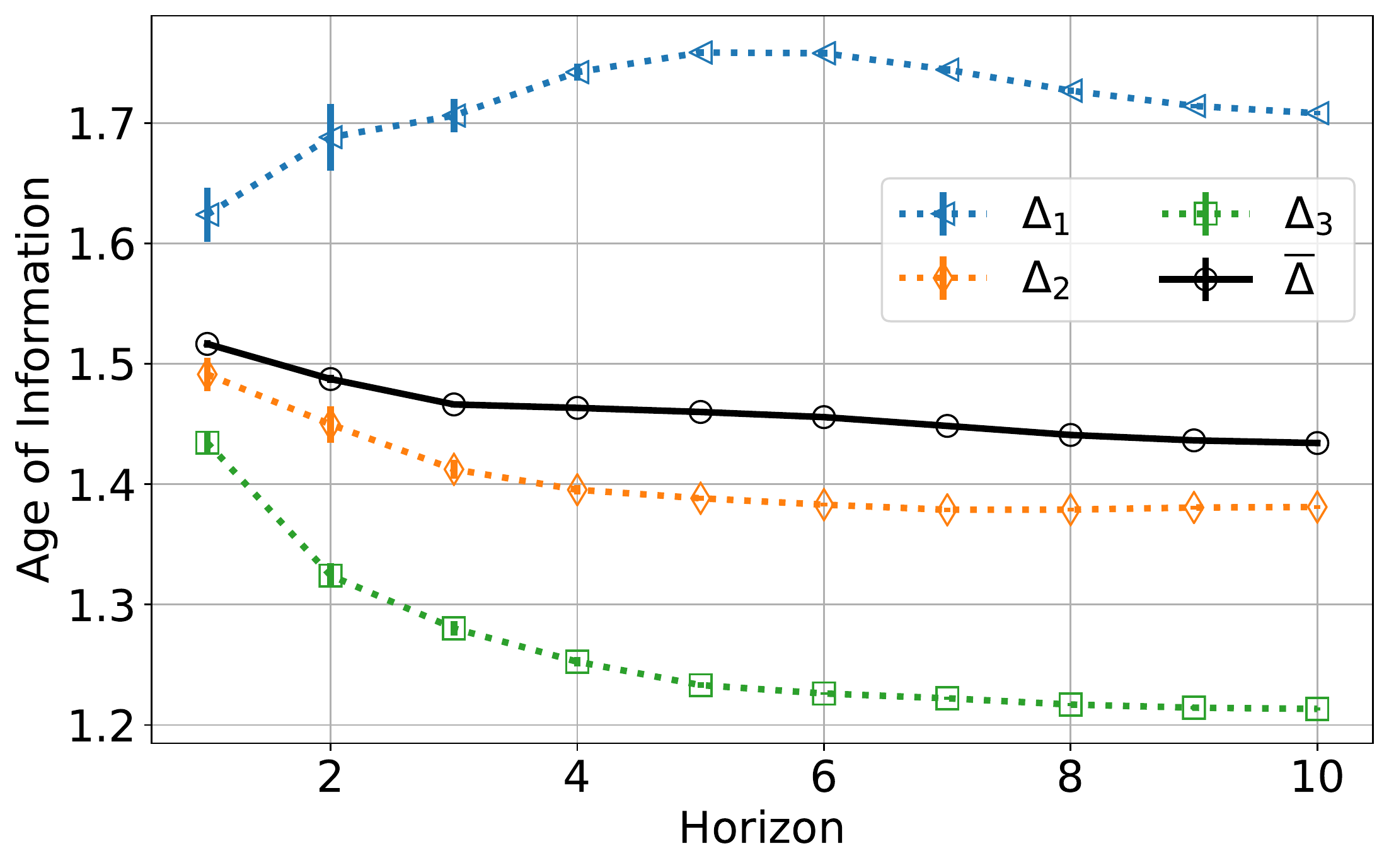}
	\caption{Solid line $\overline{\Delta}$ illustrates the average AoI in the network per time slot. Dashed lines $\Delta_1$, $\Delta_2$ and $\Delta_3$ show the the average AoI per time slot of each class with $A_1 = 1.0$, $A_2=1.25$ and $A_3=1.5$, respectively. }
	\label{fig:AoIseparate}
\end{figure}

Fig.~\ref{fig:AoIseparate} illustrates the resulting AoI performance. As we can see, the average AoI differs for each sub-system. This is an expected result, since the FH aims to minimize the short-term MSE trajectory, regardless of how often any loop is granted medium access. This leads to unequal distribution of resources due to their heterogeneous system dynamics, thus to different update frequencies. In particular, we can see that the most critical sub-system with $A_i = 1.5$ is scheduled in average more frequently, because in our scenario a lower AoI indicates a more frequent update rate. Additionally, we observe a reduction in $\Delta_2$ and $\Delta_3$ as $H$ increases. This effect results from the increasing foresight of the scheduler for possible high future costs in case of consecutive unsuccessful transmissions. Note that, although $\Delta_3$ is the lowest among all $\Delta_i$, sub-system $3$ is still the one with the highest MSE in the network.

From Fig.~\ref{fig:AoIseparate} it is evident that the FH scheduler modifies the distribution of network resources among sub-systems as a different $H$ is selected. On the other hand, Fig.~\ref{fig:MSEavg} shows that the additional performance gain it brings in terms of cost is not as obvious as the induced complexity, which grows exponentially with $H$. The effect of increasing complexity is given in table below which compares the average number of nodes in the root tree during simulations (FH) to the worst-case bound without any search space reduction: 
\begin{center}
	\begin{tabular}{ c | c |c | c | c}
		$H$ & 1 & 5 & 9 & 10 \\ 
		\hline
		FH & $4.6$ & $5.9\cdot10^2$ & $5.7\cdot10^4$ & $1.8\cdot10^5$ \\		
		WC & $5.0$ & $1.4\cdot10^3$ & $3.5\cdot10^5$& $1.4\cdot10^6$ \\ 
	\end{tabular}
\end{center}
We observe a complexity reduction up to approximately $87\%$ in our implementation compared to the worst-case scenario with $\frac{(N + 1)^{H+1} - 1}{N}$ nodes. This effect is caused by constraining the action set as in Eq.~\eqref{eq:admissibleactions} and narrowing down our search space within the class of admissible scheduling policies.

\section{Conclusion}
\label{sec:conclusion}


In this work, together with the system dependent parameters, we propose a novel AoI based control-aware scheduling algorithm that considers a finite horizon $H$ for a single-hop network with time varying channel conditions. To the best of our knowledge this is the first algorithm that guarantees optimal scheduling decisions with time varying channel conditions. We show by simulations that the quality of control (QoC) can be improved by increasing $H$ since the scheduler becomes more farsighted. On the other hand, the complexity of obtaining such an optimal policy increases exponentially with $H$. Nevertheless, we show that in spite of  the increase in complexity, the QoC does not improve in the same order. In fact, no significant improvement of QoC is observed after $H=5$ in our setup. Obviously, such a point is strongly scenario dependent and we conclude that one has to find a trade-off between optimality by increasing $H$ and complexity by decreasing $H$. {Most importantly, such a trade-off can be found easily through required modifications to take the dynamics of LTI control systems at hand into account. The proposed scheduler can be deployed in M2M and industrial IoT networks supporting control systems, where the resource management is done centrally.}

\bibliography{bibliography}
\bibliographystyle{IEEEtran}
\end{document}